\documentclass[prl,aps,superscriptaddress,amsmath,amssymb,twocolumn,floatfix,amsfonts,showpacs]{revtex4-1}
\usepackage{times}
\usepackage[varg]{txfonts}
\usepackage{textcomp}
\usepackage{graphicx}
\usepackage{subfigure}
\usepackage{tabu}
\usepackage{color}
\usepackage[colorlinks=true,citecolor=blue,urlcolor=blue,linkcolor=blue,hyperindex]{hyperref}
\usepackage{braket}
\usepackage{float}
\usepackage{overpic}
\usepackage{amssymb}

\allowdisplaybreaks

\usepackage{times}

\def\be{\begin{equation}} \def\ee{\end{equation}}
\def\bea{\begin{eqnarray}} \def\eea{\end{eqnarray}}

\def\bk{{\bf k}}

\def\be{{\bf e}}
\def\bd{{\bf d}}

\def\rw{\rightarrow}

\begin{document}

\title{Nonanalyticity of circuit complexity across topological phase transitions}

\author{Zijian Xiong}

\author{Dao-Xin Yao}\email{yaodaox@mail.sysu.edu.cn}

\author{Zhongbo Yan}\email{yanzhb5@mail.sysu.edu.cn}
\affiliation{State Key Laboratory of Optoelectronic Materials and Technologies, School of Physics, Sun Yat-Sen University, Guangzhou 510275, China}

\date{\today}
\begin{abstract}
The presence of nonanalyticity in observables is a manifestation of phase transitions. Through
the study of two paradigmatic topological models in one and two dimensions, in this work we show that
the circuit complexity based on our specific quantification
can reveal the occurrence of topological phase transitions, both in and out
of equilibrium, by the presence of nonanalyticity.
By quenching the system out of equilibrium, we find that the circuit complexity
grows linearly or quadratically in the short-time regime if
the quench is finished instantaneously or in a finite time, respectively.
Notably, we find that for both
the sudden quench and the finite-time quench,
a topological phase transition in the pre-quench Hamiltonian will be manifested by the presence
of nonanalyticity in the first-order or second-order derivative of circuit complexity with respect to
time in the short-time regime, and a topological phase transition in the post-quench Hamiltonian will be manifested by the presence of nonanalyticity in the steady value of circuit complexity in the long-time regime.
We also show that the increase of dimension does not remove, but only weakens the nonanalyticity
of circuit complexity.  Our findings can be tested in quantum simulators and cold-atom systems.

\end{abstract}

\maketitle

\section{I. Introduction}

In the development of physics, concepts from one field sometimes are
able to revolutionize the understanding of other fields. Recently, the
complexity, which was originally developed in quantum information science
to characterize how difficult to prepare one target state from certain reference state\cite{papadimitriou2003com,arora2009com}, has
been brought into the fields of holography and black hole physics\cite{susskind2016computational,Susskind2016CV,Stanford2014CV,Brown2016CAa,Brown2016CAb}. Among various progresses,
the two conjectures, namely ``complexity equals volume''\cite{susskind2016computational,Susskind2016CV,Stanford2014CV}  and ``complexity equals action''\cite{Brown2016CAa,Brown2016CAb},
have attracted particular attention and triggered active investigation of complexity in these fields\cite{Alishahiha2015com,Barbón2016com,Lehner2016action,Ben-Ami2016,Carmi2017comment,
Chapman2017com,Caputa2017com,Cano2018com,Fu2018com,Swingle2018,Couch2018,
Brown2018second,Goto2019,Brown2019JT,Bernamonti2019com,Caputa2019com,Jiang2018action,Chapman2018com},
hopefully producing new insights in understanding quantum gravity.

In the study of complexity, its quantification is a central topic\cite{Chapman2018com,Jefferson2017circuit,Hackl2018,Camargo2019,Jiang2019com}.
According to the original quantum-circuit definition,
the complexity (in this context, it is usually dubbed
circuit complexity) corresponds
to the minimum number of elementary gates required to realize a
unitary operator $U$ which transfers the reference state $|\psi_{R}\rangle$
to the target state $|\psi_{T}\rangle$, i.e., $|\psi_{T}\rangle=U|\psi_{R}\rangle$.
As the choice of elementary gates itself has a lot of freedom, the quantification based on
this principle is apparently  not an easy task. A breakthrough was made by Nielsen and collaborators
who provided a  geometric interpretation to the circuit complexity\cite{nielsen2005geometric, nielsen2006quantum,dowling2008geometry}.
Concretely, as a desired unitary operator can be generated by some time-dependent Hamiltonian
$H(t)$, they impose a cost function $F[H(t)]$ which defines a Riemannian geometry
on the space of unitary operations, then the circuit complexity is shown to
correspond to the minimal geodesic length of the Riemannian
geometry.

The minimal-geodesic-length quantification makes the circuit complexity
be a geometric quantity.  In contemporary physics, another geometric quantity
of great interest is  the so-called topological invariant which mathematically  characterizes the global
geometric property of a closed manifold. Over the past decades, this concept
has been demonstrated to play a fundamental role in characterizing new phases of matter
in condensed matter physics\cite{hasan2010,qi2011,Chiu2016RMP,Wen2017review}. As the topological invariant of a phase
is defined in terms of the wave function of ground state or the underlying Hamiltonian, a change of topological
invariant (or say topological phase transition) thus indicates a dramatic change of the geometry of
the manifold defined by the wave function of ground state or the Hamiltonian. Therefore, it is quite natural
to expect that a topological phase transition can also be manifested through the circuit complexity if the reference and target states
correspond to two distinct ground states. Very recently, Liu {\it et al.}\cite{Liu2019com} did find that
both equilibrium and dynamical topological phase transitions in the one-dimensional Kitaev model\cite{Kitaev2001unpaired} can be revealed through
the circuit complexity. Concretely, they found that by choosing the ground states of the Kitaev model as the reference
and target states, the circuit complexity exhibits nonanalytic behavior at the critical points where the target ground state undergoes a
dramatic change in topology.
In addition, when the Kitaev model undergoes a sudden quench, they found that
by choosing the pre-quench ground state as the reference state and
the post-quench unitary evolution state as the target state, the circuit complexity first increases and then saturates,
with the steady value displaying nonanalyticity when the Kiatev model is quenched across a critical point.

As the cost function itself has some arbitrariness, the quantification of circuit complexity is
not unique. It is therefore worthy to find out that whether the nonanalyticity of
circuit complexity is preserved or not when a different quantification is adopted. Given this,
in this work we consider a cost function distinct from Ref.\cite{Liu2019com} to
quantify the circuit complexity. For generality, we further consider two
topological models with  richer phase diagrams. As dimension is known to have strong impact on
both classical and quantum phase transitions, here we consider that the two topological models take different dimensions
to reveal its impact on the circuit complexity.
To reveal whether the presence of
nonanalyticity in the circuit complexity is generic or not when the system is quenched
out of equilibrium, both sudden quench and  finite-time quench are investigated.
Our main findings can be summarized as follows: (i) while
we adopt a distinct quantification, the presence of nonanalyticity in circuit complexity across
topological phase transitions holds.
(ii) We find that the long-time behavior of the post-quench circuit complexity is not sensitive
to the way of quench. Similarly to Ref.\cite{Liu2019com}, we find that for
both the sudden quench and the finite-time quench,
the steady value of circuit complexity based on our quantification in the long-time regime
can reveal the occurrence of topological phase transitions in the post-quench Hamiltonian
by the presence of nonanalyticity.
(iii) The short-time growth behavior of the post-quench circuit complexity, however,
is found to depend on the way of quench. It grows linearly/quadratically with time for
the sudden/finite-time quench in the short-time regime. Notably, we find that
the first-order (for sudden quench) or second-order (for finite-time quench)
derivative of circuit complexity with respect to time in this regime
can reveal the occurrence of topological phase transitions in the pre-quench Hamiltonian
by the presence of nonanalyticity. It is noteworthy that unlike
the steady value, the determination of the short-time growth behavior can be done
very quickly since it requires
very little information in the short-time regime,
therefore, we believe that this finding
is particularly useful for the experimental study of topological
phase transitions. (iv) We find that the increase of dimension does not remove,
but only weakens the nonanalyticity of circuit complexity, demonstrating the generality
of the underlying physics. As in recent years equilibrium and dynamical
topological phase transitions are of great interest both in theory\cite{Heyl2013dynamical,Vajna2015dynamical,Budich2016dynamical,Sharma2016,Bhattacharya2017dynamical,Wang2017quench,Yu2017quench,
Liang2018dynamical,zhang2018dynamical,heyl2018dynamical,Heyl2018dynamicalb,Qiu2018dynamical,Chang2018quench,Ezawa2018quench,Gong2018quench} and in experiment\cite{Jurcevic2017,flaschner2018observation,
tarnowski2019measuring,Sun2018dtpt,Yi2019}, our findings may shed new light on
this active field.

The paper is organized as follows. In Sec.II, we introduce our quantification of circuit complexity and
show that across various topological phase transitions in one dimension, the circuit complexity always displays nonanalyticity
at the critical points. In Sec.III, we consider that the one-dimensional generalized Kitaev model
is suddenly quenched. We study the evolution of circuit complexity after the sudden quench and show
that the short-time and long-time behavior of the post-quench circuit complexity
can respectively reveal the occurrence of topological phase transitions in the pre-quench
and post-quench Hamiltonian. In Sec.IV, we consider that the quench process is finished in
a finite time and show that the conclusions for the sudden quench are preserved.
In Sec.V, we generalize the study to two dimensions and discuss the impact of dimension on the
circuit complexity. We conclude in Sec.VI. Some calculation details are relegated to appendices.

\section{II. Circuit complexity and topological phase transitions in one dimension}

We start with a generalized Kitaev model which takes the form
$H=\frac{1}{2}\sum_{k}\psi_{k}^{\dag}H(k)\psi_{k}$ with
$\psi_{k}=(c_{k},c_{-k}^{\dag})^{T}$ and
\begin{eqnarray}
H(k)&=&(-t_{1}\cos k-t_{2}\cos 2k-\mu)\tau_{z}+(\Delta_{1}\sin k+\Delta_{2}\sin 2k)\tau_{y} \nonumber\\
&\equiv& d_{z}(k)\tau_{z}+d_{y}(k)\tau_{y},\label{oned}
\end{eqnarray}
where $\tau_{x,y,z}$ are Pauli matrices in particle-hole space,  $t_{1}$ $(\Delta_{1})$ and $t_{2}$ $(\Delta_{2})$ represents the nearest-neighbor and next-nearest-neighbor hopping (pairing) amplitude, respectively, and
$\mu$ is the chemical potential. For convenience, the lattice constant is set to unit throughout this work.

\begin{figure}[htbp]
\centering
\includegraphics[width=0.42\textwidth]{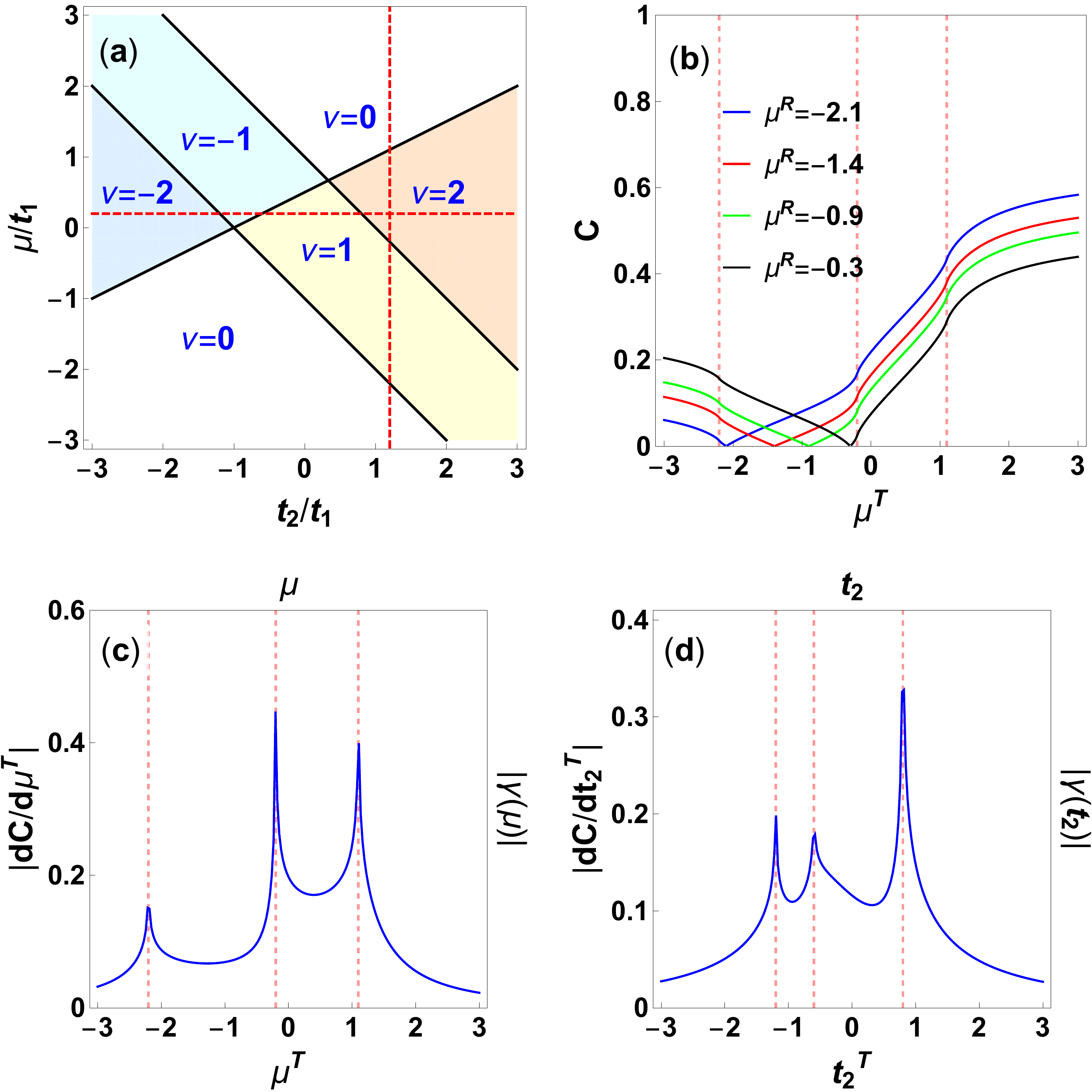}
\caption{ (Color online) (a) Phase diagram.
$\nu$ denotes the winding number, and the horizontal ($\mu/t_{1}$=0.2) and
vertical ($t_{2}/t_{1}=1.2$) red dashed lines are two paths
that we choose to vary the parameters. (b) Circuit complexity (in units of system size,
below this is implicitly assumed) for several reference states. Parameters are
$t_{1}=1$, $t_{2}=1.2$, and $\Delta_{1}=\Delta_{2}=1$. (c) The derivative of  circuit complexity with respect to $\mu^{T}$.
Parameters are the same as in (b).  $|dC/d\mu^{T}|$ turns out to be independent of $\mu^{R}$,
and is equal to $\gamma(\mu)$. (d) The derivative of circuit complexity with respect to $t_{2}^{T}$.
Parameters are $t_{1}=1$, $\mu=0.2$, and $\Delta_{1}=\Delta_{2}=1$. $|dC/dt_{2}^{T}|$ is also found to be independent of $t_{2}^{R}$
and coincide with $\gamma(t_{2})$. Throughout this work,
vertical pink dashed lines (see (b)(c)(d)) correspond to the critical points at which topological
 phase transitions take place.}\label{fig1}
\end{figure}

Owing to the presence of chiral symmetry, i.e., $\{\tau_{x},H(k)\}=0$, this model belongs to
the class BDI\cite{schnyder2008,kitaev2009,ryu2010} and is known to be characterized by the winding number defined as\cite{ryu2010}
\begin{eqnarray}
\nu=\frac{i}{4\pi}\int_{-\pi}^{\pi}dk \text{Tr}[\tau_{x}H^{-1}(k)\partial_{k}H(k)].\label{winding}
\end{eqnarray}
According to the above formula, we present the phase diagram corresponding to
$\Delta_{1}=\Delta_{2}\neq0$ in Fig.\ref{fig1}(a). Noteworthily, as a
topological phase transition is associated with the closure of bulk energy gap,
the phase boundaries in the phase diagram can be easily determined, which
are found to be the lines satisfying $\mu=t_{1}-t_{2}$, $\mu=-t_{1}-t_{2}$ and $\mu=(t_{1}+t_{2})/2$.
In comparison to the standard Kitaev model which only involves
the nearest-neighbor hopping and pairing \cite{Kitaev2001unpaired}, it is readily seen that
the presence of additional next-nearest-neighbor hopping and pairing leads
to a richer phase diagram, and therefore more topological phase transitions
can be investigated to demonstrate the generality of the presence of nonanalyticity
in circuit complexity.

The generalized Kitaev model above describes a spinless superconductor.
By following the  standard Bogoliubov transformation, the model can be diagonalized
and accordingly the ground-state wave function can be obtained, which reads\cite{SM}
\begin{eqnarray}
|\Omega\rangle=\prod_{k>0}|\psi_{k}\rangle
=\prod_{k>0}(\cos(\theta_{k}/2)+i\sin(\theta_{k}/2)c_{k}^{\dag}c_{-k}^{\dag})|0\rangle,\label{GS}
\end{eqnarray}
where $\theta_{k}=\arctan (d_{y}(k)/d_{z}(k))$. As here the momentum is a good quantum number,
we can treat each $k$ independently\cite{Liu2019com}. For each $k$, one can see that the state is a superposition
of $|0\rangle$ and $c_{k}^{\dag}c_{-k}^{\dag}|0\rangle$, and the superposition
is characterized by a single parameter $\theta_{k}/2$ (noteworthily, $\theta_{k}/2$ and
$\theta_{k}/2+\pi$ are equivalent as the $\pi$ difference only results in a global phase
difference to the wave function).
Taking the ground state with $\theta_{k}^{R}/2$ and $\theta_{k}^{T}/2$  as the reference
state $|\psi_{k}^{R}\rangle$ and target state $|\psi_{k}^{T}\rangle$, respectively,  then the effect of
unitary operator is equivalent to doing a transport from the point $\theta_{k}^{R}/2$ to the point $\theta_{k}^{T}/2$
on the unit circle. Apparently,  on the circle the length of the arc connecting $\theta_{k}^{R}/2$ and $\theta_{k}^{T}/2$
provides a natural measure of the distance between the two states. Owing to the equivalence
between $\theta_{k}/2$ and $\theta_{k}/2+\pi$, the minimal length of the arc is in fact given by
$\arccos|\langle \psi_{k}^{T}|\psi_{k}^{R}\rangle|$, where $|\langle \psi_{k}^{T}|\psi_{k}^{R}\rangle|$
is known as fidelity. In fact, such a quantification  corresponds to
the well-known quantification of circuit complexity in terms of
inner-product metric\cite{Brown2019com,SM}.
Throughout this work, we adopt this quantification
which has a simple and clear geometric interpretation.

Accordingly, if we start with a ground state $|\Omega^{R}\rangle$ and adiabatically transfer it to
another ground state $|\Omega^{T}\rangle$,
then the corresponding circuit complexity is
\begin{eqnarray}
C=\sum_{k>0}\arccos|\langle \psi_{k}^{T}|\psi_{k}^{R}\rangle|
=\sum_{k>0}\arccos|\cos(\Delta\theta_{k}/2)|,\label{onecc}
\end{eqnarray}
where $\Delta\theta_{k}=\theta_{k}^{T}-\theta_{k}^{R}$. For each $k$, one can see that the maximum
value is $\pi/2$, which corresponds to that $|\psi_{k}^{R}\rangle$ and $| \psi_{k}^{T}\rangle$ are orthogonal.
For comparison, the formula for circuit complexity in Ref.\cite{Liu2019com} takes the form $C=\sum_{k}(\Delta\theta_{k}/2)^{2}$.
While in essence the two formulas only differ by a square-root operation, in the following one will see
that remarkable differences will arise in the behaviors of circuit complexity.  To simplify the analysis of circuit complexity,
below we will restrict ourselves to ground state evolutions
corresponding to the variation of only one parameter of the model. Concretely, we
will fix $(t_{1}, \Delta_{1},\Delta_{2})$ and only vary either $\mu$ or $t_{2}$.

In Fig.\ref{fig1}(b), we present the circuit complexity
associated with the variation of $\mu$. The result clearly demonstrates the presence of nonanalyticity
in the $C$-$\mu^{T}$ curve. The nonanalytic behavior becomes even more apparent by
performing a first-order derivative, i.e., $dC/d\mu^{T}$.
As shown in  Fig.\ref{fig1}(c), divergence appears exactly at the critical
points of topological phase transitions. Furthermore, it is readily found that in Fig.\ref{fig1}(c)
$dC/d\mu^{T}$ does not depend on $\mu^{R}$, that is, the first-order derivative of
circuit complexity does not depend on the choice of reference state. Noteworthily, such an independence
of reference state is absent in Ref.\cite{Liu2019com}, indicating that
different quantifications  may lead to remarkable different behaviors in the circuit complexity.

Let us now give a simple explanation of the presence of nonanalyticity. For this generalized
Kitaev model, its topological invariant, the winding number, characterizes the number of times that the
vector $\bd(\bk)=(d_{y}(\bk),d_{z}(\bk))$ winds around the origin when $k$ goes from $-\pi$ to $\pi$.
When a topological phase transition occurs, the number of times that the vector $\bd(\bk)$
winds around the origin has a discrete change. Accordingly, the angle $\theta_{k}$, which characterizes
the orientation of the vector $\bd(\bk)$,
must jump at some $k$. Apparently, $\Delta \theta_{k}$ for these $k$ also jump at the critical points.
Then according to Eq.(\ref{onecc}), it is readily seen
that the circuit complexity will be nonanalytic at the critical points.

Before proceeding, here we define a  quantity,
\begin{eqnarray}
\gamma(\lambda)\equiv \sum_{k>0}\lim_{\delta\lambda\rw0^{+}}\frac{\arccos|\langle \psi_{k}(\lambda+\delta\lambda)|\psi_{k}(\lambda)\rangle|}{\delta\lambda},
\end{eqnarray}
where $\lambda$ denotes some  parameter of the Hamiltonian. The physical meaning
of this quantity is quite obvious. It  characterizes the growth rate of circuit complexity
when a state is adiabatically transferred to its nearby state.  For the sake of discussion
convenience, we name
it {\it adiabatical growth rate}. According to Eq.(\ref{GS}), a short calculation reveals\cite{SM}
\begin{eqnarray}
\gamma(\lambda)=\sum_{k>0}\frac{1}{2}|\frac{d\theta_{k}}{d\lambda}|.\label{agr}
\end{eqnarray}
As mentioned above, when a topological phase transition occurs, $\Delta \theta_{k}$
will dramatically change for some $k$. Therefore, it is quite obvious that $\gamma(\lambda)$
is also nonanalytic at the critical points. Interestingly,  we find that $|dC/d\mu^{T}|$ and $\gamma(\mu)$
coincide with each other, as shown in Fig.\ref{fig1}(c). The divergence of $\gamma(\mu)$ at
the critical points indicates
that the adiabatical growth rate goes
divergent when the state gets close to a critical point. In other words,
when the reference state gets more close to a critical point, it becomes
more difficult to prepare the target state in an adiabatical way.
Noteworthily, the overlap of the adiabatical growth rate and the first-order derivative  is not accidental. As shown in
Fig.\ref{fig1}(d), it also appears when we fix $\mu$ and vary $t_{2}$. This overlap is apparently related
to the result that the first-order derivative of circuit complexity is independent of the reference state, indicating that
this property is tied to our quantification.

\begin{figure}[t]
\centering
{\subfigure{\includegraphics[width=0.4\textwidth]{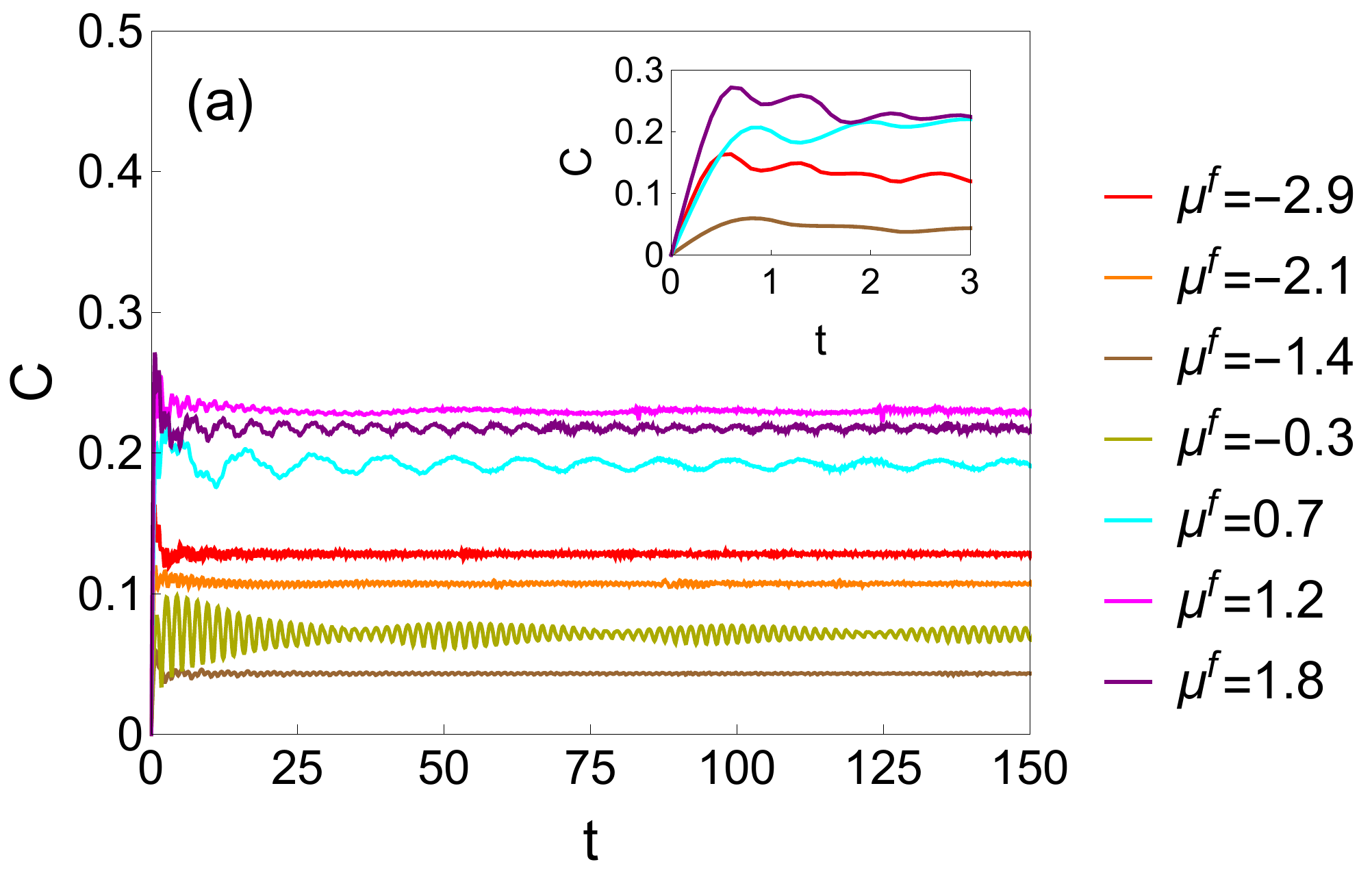}}}
{\subfigure{\includegraphics[width=0.2\textwidth]{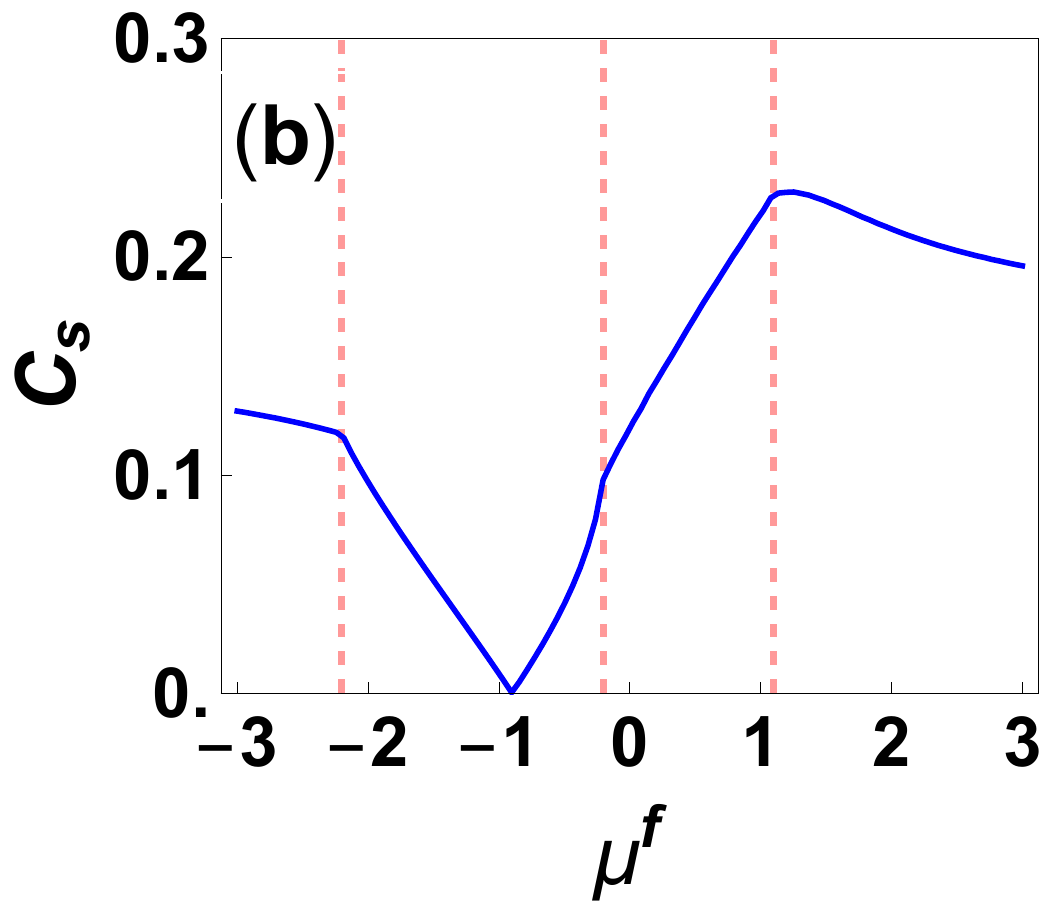}}}
{\subfigure{\includegraphics[width=0.2\textwidth]{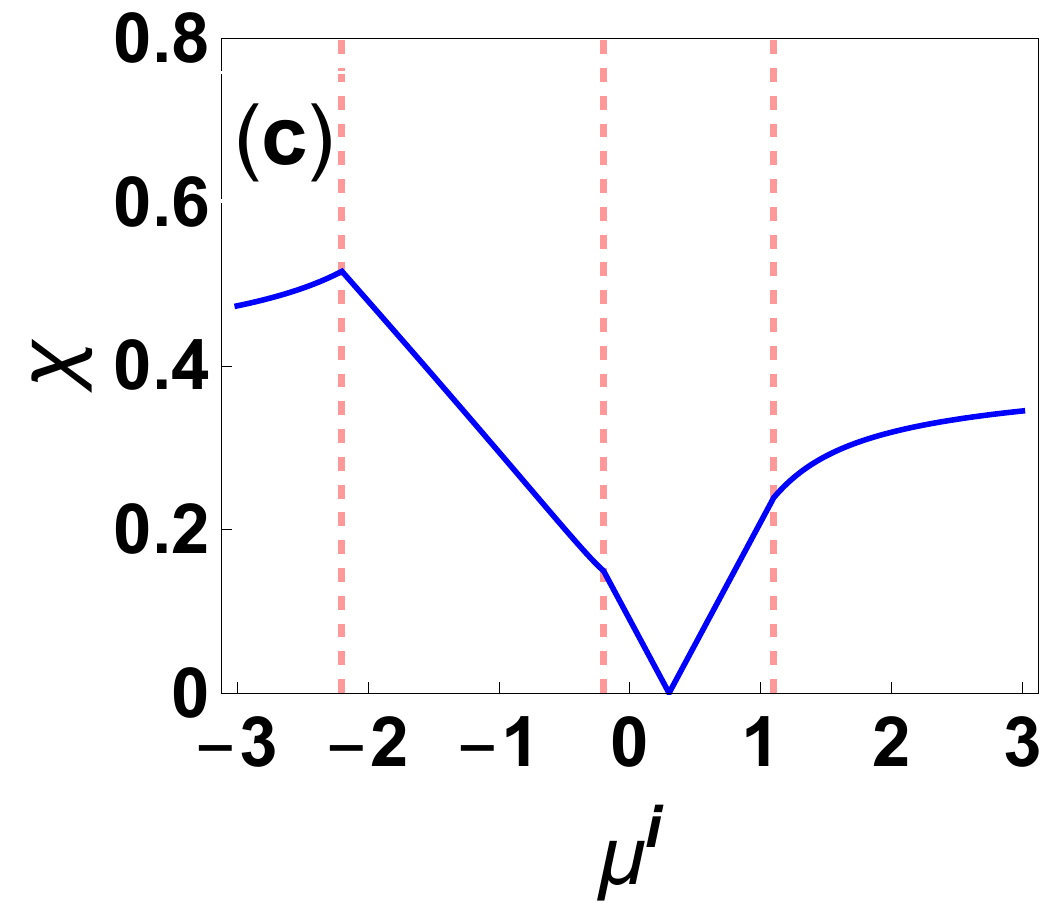}}}
\caption{ (Color online) Common parameters are $t_{1}=1$, $t_{2}=1.2$, and $\Delta_{1}=\Delta_{2}=1$. (a) The evolution of post-quench circuit complexity.  $\mu^{i}=-0.9$.  (b) Steady value of post-quench circuit complexity shown in (a). (c) Dynamical growth rate. $\mu^{f}=0.3$ is fixed. }\label{fig2}
\end{figure}

\section{III. Sudden quench and circuit complexity evolution}

In equilibrium, as the Hamiltonian
and the ground state are tied to each other, the topological invariants defined in terms of
them are equivalent. When out of equilibrium,
however, as the underlying instantaneous wave function in general does not correspond to the ground state of
the instantaneous Hamiltonian, the topological invariants defined in terms of instantaneous Hamiltonian (labeled as $\nu_{H}$) and
instantaneous wave function (labeled as $\nu_{w}$) are not guaranteed to be equivalent. In particular, when a system is isolated
from the environment, the wave function will follow unitary evolution, and so $\nu_{w}$ will keep its value
no matter how  $\nu_{H}$ changes\cite{Caio2015quench,Luca2015quench}.
Therefore, for an isolated system out of equilibrium, a topological phase transition
can only be defined as a change of $\nu_{H}$. As out of equilibrium, transitions associated
with a change of $\nu_{H}$ are usually dubbed dynamical topological phase transitions.
In the following, we focus on such isolated systems and first
investigate the evolution of
circuit complexity after a sudden quench\cite{Alves2018,Jiang2018quench,Fan2018,Ali2018quenchb,Ali2019quench,Moosa2018quench,Camargo2019quench,Ageev2019quench}.

For concreteness, we consider that for $t<0$, the system is described by $H_{i}$ and stays at its ground state $|\Omega^{i}\rangle$.
At $t=0$, the  Hamiltonian is suddenly quenched to $H_{f}$, and afterwards it keeps as $H_{f}$. Accordingly, the wave function
at $t>0$ is given by $|\Omega(t)\rangle=e^{-iH_{f}t}|\Omega^{i}\rangle$. For each $k$, we have
$|\psi_{k}(t)\rangle=e^{-iH_{f}(k)t}|\psi_{k}^{i}\rangle$. Thus, the post-quench circuit complexity
is given by $C(t)=\sum_{k>0}\arccos|\langle \psi_{k}(t)|\psi_{k}^{i}\rangle|$.
A short calculation reveals\cite{SM}
\begin{eqnarray}
C(t)=\sum_{k>0}\arccos\sqrt{1-\sin^{2}\Delta\tilde{\theta}_{k}\sin^{2} (E_{f}(k)t)},
\end{eqnarray}
where $\Delta\tilde{\theta}=\theta_{k}^{f}-\theta_{k}^{i}$ with $\theta_{k}^{f(i)}=\arctan (d_{y}^{f(i)}(k)/d_{z}^{f(i)}(k))$,
and $E_{f}(k)=\sqrt{(d_{y}^{f}(k))^{2}+(d_{z}^{f}(k))^{2}}$.

By fixing $\mu_{i}$, we show the evolution of post-quench circuit complexity for a series of $\mu_{f}$ in Fig.\ref{fig2}(a). One can see
 that after the quench, the circuit complexity will first grow linearly with time (see the inset)
and then saturate  with some degree of oscillations. While adopting  a different quantification,
we find that similarly to Ref.\cite{Liu2019com}, here the steady value in the long-time regime also
exhibits nonanalyticity at the critical points where the topology of the post-quench Hamiltonian
$H_{f}$ undergoes a change, as shown in Fig.\ref{fig2}(b).
Noteworthily, as oscillations always appear, we take the average of $C(t)$ over a sufficiently long time as
the steady value. Throughout this work we take $C_{s}=(\int_{t_{i}}^{t_{f}}C(t)dt)/(t_{f}-t_{i})$, with
$t_{i}=50$ and $t_{f}=130$.

From above it is readily seen that obtaining the steady value requires us to know the evolution of $C(t)$ for quite a long time,
therefore, using the steady value to detect the occurrence of
topological phase transitions is not very efficient.
Surprisingly,  we find that the growth rate of post-quench circuit complexity
within the linear growth regime in fact can also reveal the occurrence of topological phase transitions.
In order to distinguish from the equilibrium case,
here we name it {\it dynamical growth rate}.  As the circuit complexity grows linearly right after the quench,
the dynamical growth rate is thus simply given by
\begin{eqnarray}
\chi\equiv\lim_{\delta t\rw 0^{+}}\frac{C(\delta t)}{\delta t}=\sum_{k>0}|\sin\Delta\tilde{\theta}_{k}|E_{f}(k),\label{gr}
\end{eqnarray}
or equivalently, $\chi\equiv\partial_{t}C|_{t\rw0^{+}}$. One can see that for each $k$, the dynamical growth rate is proportional to the energy. If starting from different pre-quench Hamiltonian $H_{i}$
and quenching to the same $H_{f}$ (e.g.,  $H_{f}$ can be chosen to describe a trivial phase
for which hopping and pairing are turned off, i.e., $H_{f}=-\mu\tau_{z}$),
then as $\Delta\tilde{\theta}_{k}$  displays
distinct winding behavior for $H_{i}$ with distinct $v_{H}$, $\chi$ will exhibit
nonanalyticity at the critical points where
the topology of the pre-quench $H_{i}$ undergoes a change.
In Fig.\ref{fig2}(c), we present the result for the case with a fixed $\mu^{f}$.
It is readily seen that nonanalyticity does appear at every critical points.

According to Eq.(\ref{gr}), it is quite obvious that the determination of  dynamical growth rate requires very little information
right after the quench, therefore using it to reveal the occurrence of
topological phase transitions is
much more convenient and accurate than
using the steady value suggested in Ref.\cite{Liu2019com}.
As a final remark of this section, we point out that the linear growth behavior right after
the quench is also specific to our quantification. For comparison, in Ref.\cite{Liu2019com}
the circuit complexity does not grow with time in a linear way
right after the quench. For their quantification, the slope of the circuit complexity
in the extremely short-time regime is time-dependent, and therefore it is hard to give a sharp definition
of dynamical growth rate as here, suggesting that a good choice of
the quantification may benefit its application.

\section{IV. Finite-time quench and circuit complexity evolution}

\begin{figure}[t]
\centering
\includegraphics[width=0.45\textwidth]{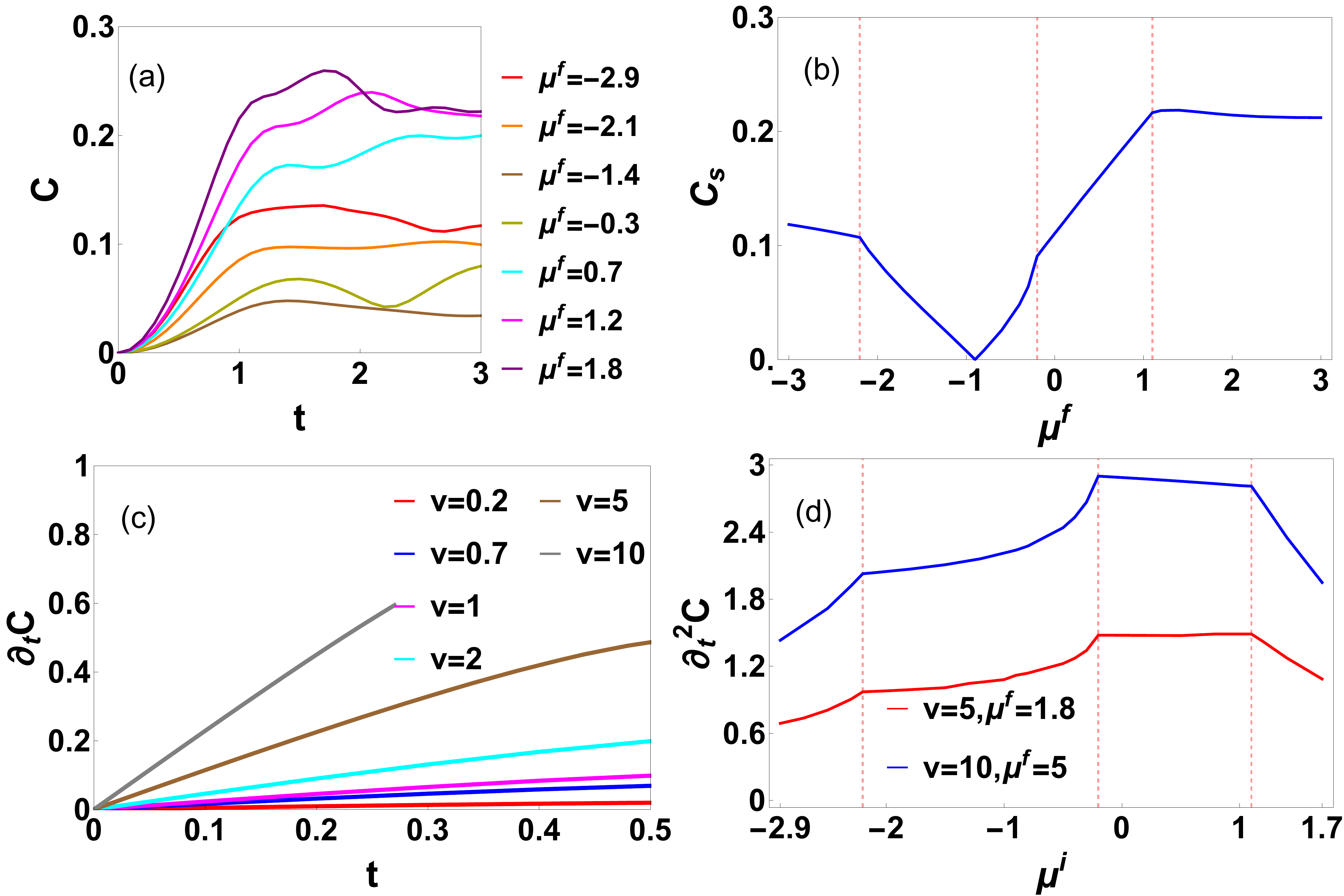}
\caption{(a) (Color online) Common parameters are $t_{1}=1$, $t_{2}=1.2$, and $\Delta_{1}=\Delta_{2}=1$. (a) The evolution of circuit complexity under a finite-time quench. The quench duration time $t_{q}$ is fixed to $1$ and $\mu^{i}=-0.9$.  (b) The dependence of steady value on $\mu^{f}$, $t_{q}$ and $\mu^{i}$ keep the same
as in (a). (c) $\partial_{t}C$, the dynamical growth rate of circuit complexity,
in the short-time regime, $\mu^{i}=-0.9$, $\mu^{f}=1.8$. The slope of $\partial_{t}C$
is found to increase with the quench speed. (d) $\partial_{t}^{2}C$ in the limit $t\rightarrow 0^{+}$.
The concrete value of $\partial_{t}^{2}C$ depends on the quench speed, but the presence of
nonanalyticity at the critical points does not rely on the quench speed.}\label{finite}
\end{figure}

For the sudden quench, the
change from the initial Hamiltonian to the final
Hamiltonian is completed instantaneously. In this section
we consider finite-time quench for which one of the parameters
in the Hamiltonian changes from
its initial value at $t\leq0$ to the final value
with a finite speed.
To be specific, we focus on the chemical potential and take
$\mu(t)=\mu^{i}+vt\Theta(t_{q}-t)
+(\mu^{f}-\mu^{i})\Theta(t-t_{q})$, where
$t_{q}=|\mu^{f}-\mu^{i}|/|v|$ is the quench duration time 
and $\Theta(x)$ is
the step function with $\Theta(x)=1/0$ for $x>/<0$.
This expression means that the chemical potential
is quenched from $\mu^{i}$ to $\mu^{f}$ with a
finite speed $v$. When $v$ goes to infinity,
it returns to the sudden quench addressed in
the previous section.

As the Hamiltonian changes with time in the domain
$0<t<t_{q}$, the circuit complexity in this domain
is given by
\begin{eqnarray}
C(t)=\sum_{k>0}\arccos |\langle \psi_{k}^{i} |Te^{-i\int_{0}^{t}H(k,t')dt'}|\psi_{k}^{i}\rangle|,
\end{eqnarray}
where $T$ denotes the time ordering operator. When $t>t_{q}$,
\begin{eqnarray}
C(t)=\sum_{k>0}\arccos |\langle \psi_{k}^{i} |e^{-iH_{f}(k)(t-t_{f})}Te^{-i\int_{0}^{t_{f}}H(k,t')dt'}|\psi_{k}^{i}\rangle|.
\end{eqnarray}

By fixing $\mu^{i}$ and $t_{q}$, we show the evolution of circuit complexity
for a series of $\mu^{f}$ in Fig.\ref{finite}(a). Similarly to the sudden-quench
case, we find that the circuit complexity will saturate with some degree of oscillations
in the long-time regime (not shown explicitly) and its steady value $C_{s}$
also displays nonanalyticity at the critical points where the topology of the post-quench Hamiltonian
undergoes a change,  as shown in Fig.\ref{finite}(b).
The similarity in the long-time regime between the sudden quench and the finite-time quench
is naturally expected since after $t_{q}$ the dynamics of the whole system
is governed by the same time-independent Hamiltonian $H_{f}$ for both cases.
When looking at the short-time regime in Fig.\ref{finite}(a),
we find that the circuit complexity does not grow linearly
as in the sudden-quench case. Instead, it grows quadratically with time
right after the start of the quench. To better show the quadratical growth,
we present the evolution of $\partial_{t} C$ in Fig.\ref{finite}(c).
It is readily seen that right after the start of the quench, $\partial_{t} C$
grows linearly with time like the circuit complexity after a sudden quench.
From Fig.\ref{finite}(c), one can also find that while $\partial_{t} C$
depends on the quench speed, its linear growth behavior
is robust against the variation of the quench speed. The linear growth behavior
of $\partial_{t} C$ leads us to expect that $\partial_{t}^{2} C$,
which can be understood as the acceleration of the circuit complexity,
will display nonanalyticity at the critical points where the topology of the pre-quench
Hamiltonian undergoes a change.  The numerical results confirm
the above expectation, as shown in Fig.\ref{finite}(d).  Therefore,
for the finite-time quench, the short-time behavior
and the long-time behavior of the circuit complexity can also respectively
reveal the occurrence of topological phase transitions in the pre-quench Hamiltonian
and in the post-quench Hamiltonian.

\section{V. Circuit complexity and topological phase transitions in two dimensions}

\begin{figure}[t]
\centering
\includegraphics[width=0.42\textwidth]{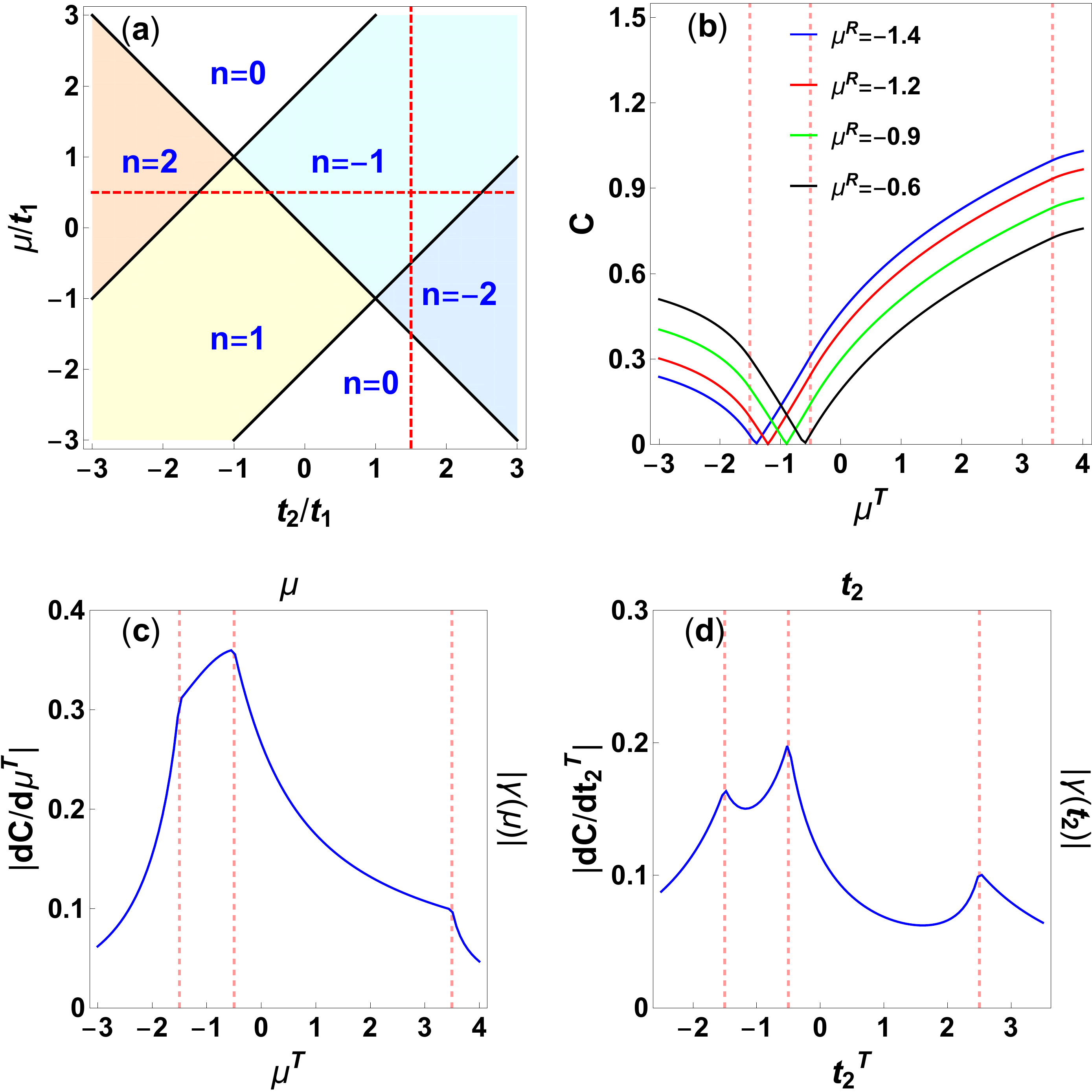}
\caption{(a) Phase diagram with $n$ denoting the Chern number. The horizontal ($\mu/t_{1}=0.5$) and vertical ($t_2/t_{1}=1.5$) red dashed lines are two paths
that we choose to vary the parameters. (b) Circuit complexity for several reference states. Parameters are
$t_{1}=1.0$ and $t_{2}=1.5$. (c) The derivative of circuit complexity with respect to $\mu^{T}$.
Parameters are the same as in (b).  (d) The derivative of circuit complexity with respect to $t_{2}^{T}$.
Parameters are  $t_{1}=1$ and $\mu=0.5$. }\label{fig4}
\end{figure}

Since a topological phase transition is always associated with a dramatic change
of the global geometric property of the underlying ground state wave function or
Hamiltonian, it is natural to expect that the circuit complexity will also exhibit
nonanalyticity in higher dimensions. To demonstrate this explicitly, here we consider
a paradigmatic ``$\bd\cdot\mathbf{\tau}$'' model in two dimensions for concreteness.
The model reads $H=\frac{1}{2}\sum_{\bk}\psi_{\bk}^{\dag}H(\bk)\psi_{\bk}$, with $H(\bk)=\bd(\bk)\cdot\mathbf{\tau}$
and \cite{Ezawa2018quench}
\begin{eqnarray}
d_{x}(\bk)&=&\sin k_{x}, \quad d_{y}=-\sin k_{y},\nonumber\\
d_{z}(\bk)&=&t_{1}(\cos k_{x} + \cos k_{y}) + t_{2} \cos k_{x} \cos k_{y}-\mu,
\end{eqnarray}
where the $d_{x}$ and $d_{y}$ terms together describe a chiral $p$-wave pairing,
and the $d_{z}$ term is the kinetic energy of normal state. As for this Hamiltonian only particle-hole
symmetry is preserved, it belongs to the class D that in two dimensions is characterized by
the first-class Chern number,
\begin{eqnarray}
n=\frac{1}{4\pi}\int_{BZ} d^{2}k \hat{\bd}(\bk)\cdot(\partial_{k_{x}}\hat{\bd}(\bk)\times\partial_{k_{y}}\hat{\bd}(\bk)),
\end{eqnarray}
where $\hat{\bd}=(d_{x},d_{y},d_{z})/\sqrt{d_{x}^{2}+d_{y}^{2}+d_{z}^{2}}$.
The geometric meaning of this formula also has a winding interpretation:
it counts the number of times that the three-component vector
$\bd(\bk)$ winds around the origin when the momentum transverses the whole first Brillouin
zone. It indicates that when a topological phase transition takes place, the orientation of the vector
$\bd(\bk)$ will also change dramatically at some momenta.
By using the above formula, we present the phase diagram in Fig.\ref{fig4}(a).

\begin{figure}[t]
\centering
{\subfigure{\includegraphics[width=0.4\textwidth]{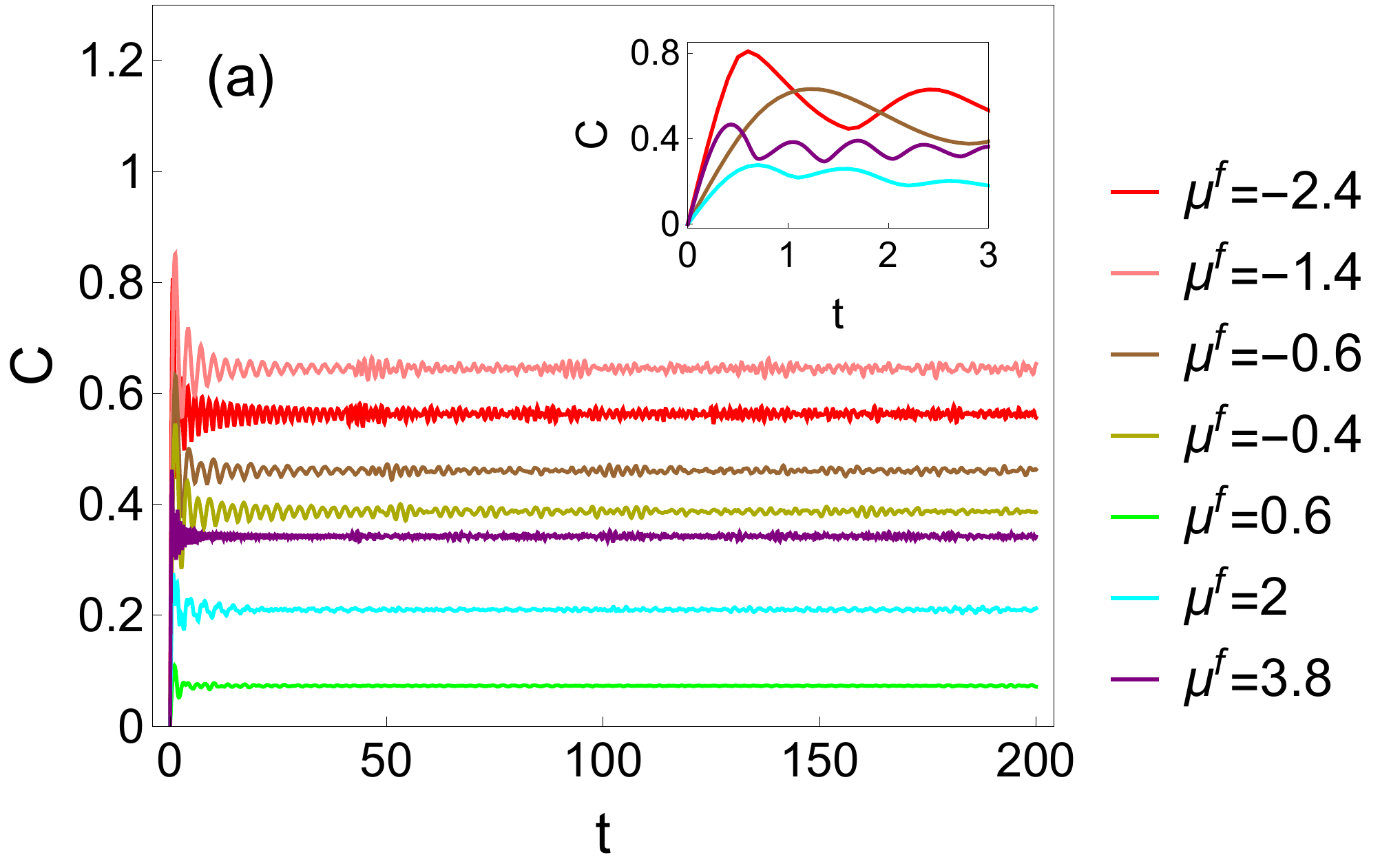}}}
{\subfigure{\includegraphics[width=0.2\textwidth]{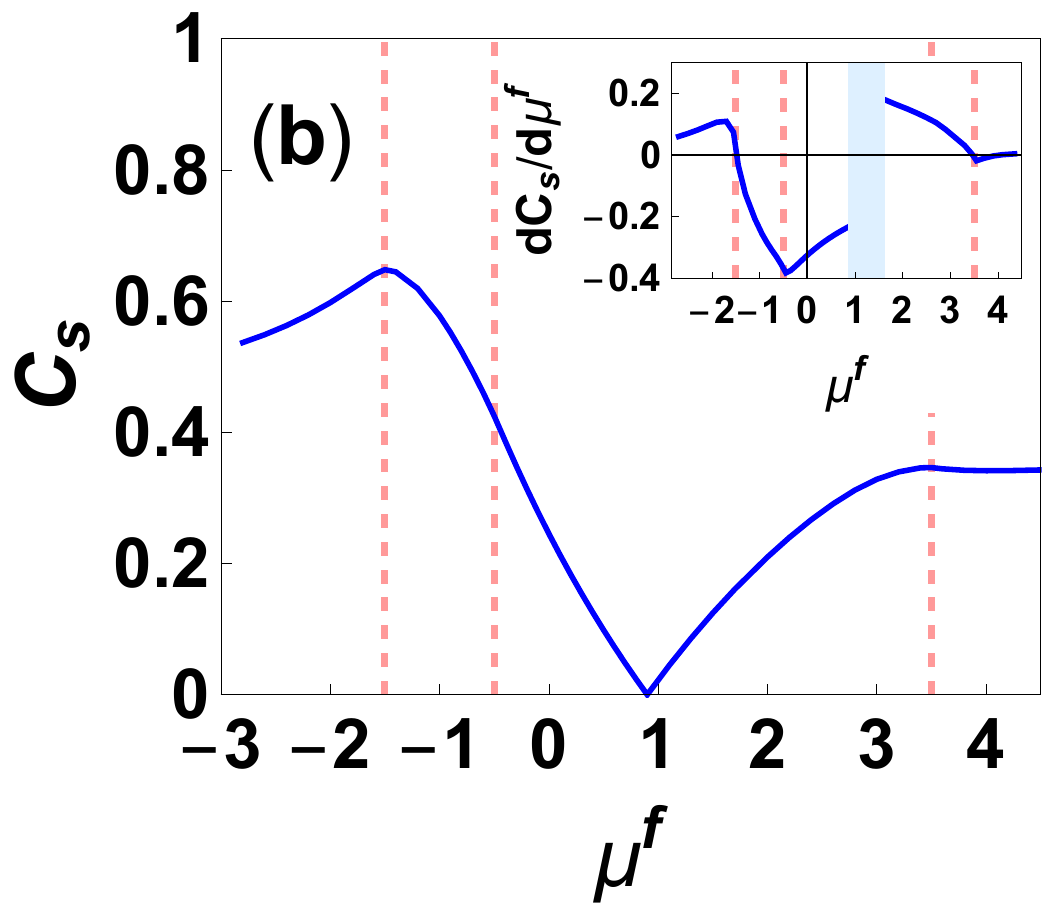}}}
{\subfigure{\includegraphics[width=0.2\textwidth]{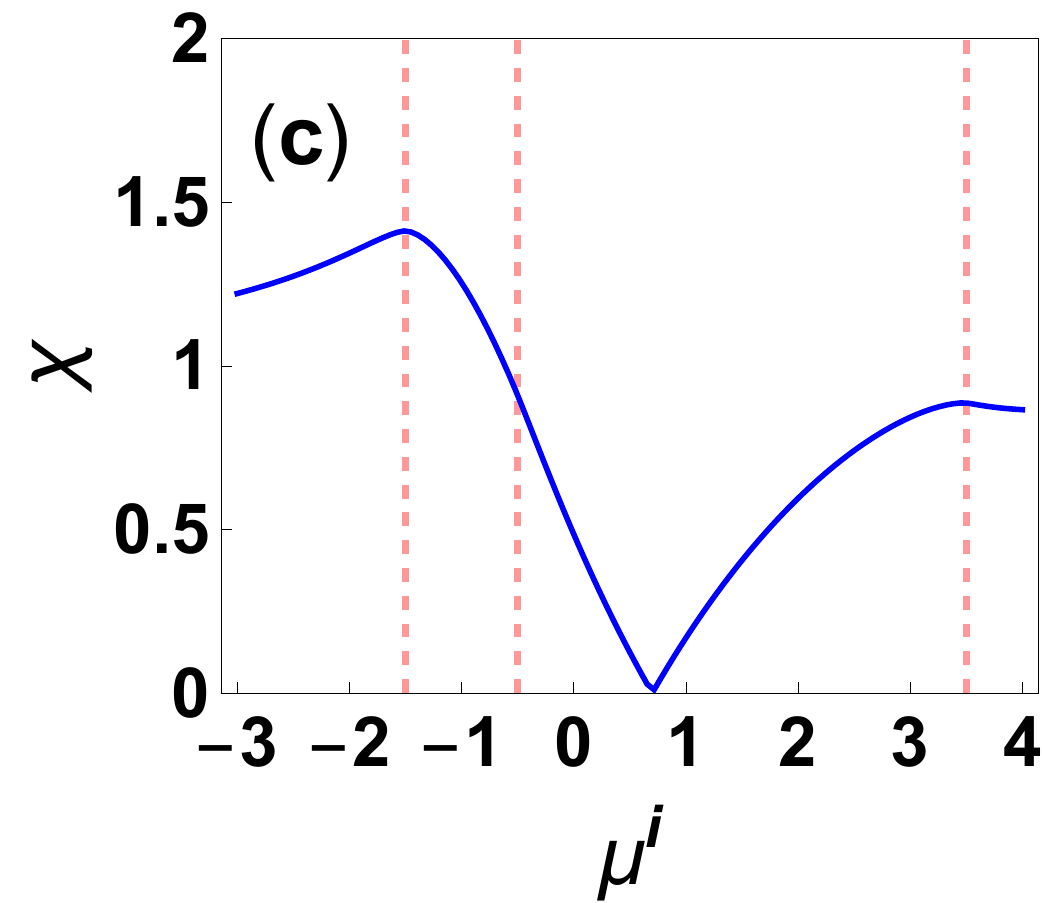}}}
\caption{ Common parameters are $t_{1}=1$, $t_{2}=1.5$. (a) The evolution of post-quench circuit complexity.  $\mu^{i}=0.9$.  (b) Steady value of post-quench circuit complexity shown in (a). (c) Dynamical growth rate. $\mu^{f}=0.7$ is fixed. }\label{fig5}
\end{figure}

Following the steps in one dimension,
we find that for an adiabatical evolution of ground states, the circuit complexity is
\begin{eqnarray}
C=\sum_{k_{x}>0,k_{y}}\arccos|\cos(\Delta\theta_{k}/2)|,\label{twocc}
\end{eqnarray}
where $\Delta\theta_{k}$ takes the same definition as in Eq.(\ref{onecc}), but here
$\theta_{k}^{T(R)}=\arctan (\sqrt{(d_{x}^{T(R)}(\bk))^{2}+(d_{y}^{T(R)}(\bk))^{2}}/d_{z}^{T(R)}(\bk))$.
 As for this model all topological
phase transitions occur at the time-reversal invariant momenta,
including $(k_{x},k_{y})=(0,0)$, $(0,\pi)$, $(\pi,0)$, and $(\pi,\pi)$, at which
$d_{x}(\bk)$ and $d_{y}(\bk)$ vanish identically, it is readily seen
that across a critical point, the orientation of the vector $\bd(\bk)$  will
reverse its direction at the respective momenta. Then according to Eq.(\ref{twocc}),
it is clear that nonanalyticity will also be present in the circuit complexity.

Similarly to the one-dimensional model, in the following we also consider that
only $\mu$ and $t_{2}$ are variables. In Fig.\ref{fig4}(b), we fix $t_{2}$ and present several $C$-$\mu^{T}$ curves corresponding to different $\mu^{R}$. In comparison to Fig.\ref{fig1}(b), it is readily seen that the increase of dimension enhances the smoothness of the $C$-$\mu^{T}$ curve.
Nevertheless, it is as expected that the circuit complexity also exhibits  nonanalyticity at all critical points, and
this feature becomes quite obvious  after performing a first-order derivative, as
shown in Fig.\ref{fig4}(c), where one can see that the overlap of $|dC/d\mu^{T}|$ and $\gamma(\mu)$ also holds.
If fixing $\mu$ and varying $t_{2}$, the results also demonstrate the presence of nonanalyticity at the critical points and the overlap of
$|dC/dt_{2}^{T}|$ and $\gamma(t_{2})$, as shown in Fig.\ref{fig4}(d).

The evolution of  circuit complexity after a sudden quench is presented in Fig.\ref{fig5}(a).
It is readily seen that the long-time and short-time behaviors are quite
similar to those in one dimension. Due to this similarity, here we will not discuss
the more complicated finite-time quench.
From Fig.\ref{fig5}(b)(c),  one can find that owing to the increase of dimension,
the $C_{s}$-$\mu^{f}$ curve, as well as
the $\chi$-$\mu^{i}$ curve both become somewhat smoother. Nevertheless, the nonanalyticity
at the critical points holds and can be revealed by performing a first-order derivative (see the inset in Fig.\ref{fig5}(b)).

Through Fig.\ref{fig4} and Fig.\ref{fig5}, we demonstrate that the circuit complexity can also reveal
 the occurrence of  topological phase transitions in higher dimensions. However, as the increase of dimension is shown to weaken the nonanalyticity,
it indicates that the application of this quantity to detect higher dimensional topological phase transitions
requires a higher precision of measurements.

\section{VI. Conclusions}

In this work, we have adopted a simple quantification of the circuit complexity distinct from Ref.\cite{Liu2019com} and
demonstrated that it can reveal the occurrence of topological phase transitions by
the presence of nonanalyticity. By studying the evolution of circuit complexity after a
quench, either sudden or finite-time, we find that the
growth behavior in the short-time regime and
the steady value in the long-time regime can respectively
reveal the occurrence of
topological phase transitions in the pre-quench Hamiltonian
and in the post-quench Hamiltonian by the presence of nonanalyticity.
Since the growth behavior in the short-time regime can be determined by very
 few measurements, its nonanalytic behavior is expected to be
very useful for the experimental
study of topological phase transitions.
Furthermore, we have investigated the effect of dimension and shown that the increase of dimension does not remove, but only weakens the nonanalyticity.
As the circuit complexity based on our quantification can be determined by measuring the fidelity, the test of our predictions
is well with the current experimental accessibility. For instance,
the two-band models studied can be simulated in a superconducting qubit, and
the evolution of states can be measured by performing  quantum state tomography\cite{Roushan2014}.
In light of the active study of equilibrium and
dynamical topological phase transitions
in quantum simulators and cold-atom systems\cite{Jurcevic2017,flaschner2018observation,
tarnowski2019measuring,Sun2018dtpt,Yi2019}, our predictions can also be tested in these platforms,
hopefully providing some new insights into understanding  various phase transitions.

\section{Acknowlegements}

 We would like to acknowledge helpful discussions with Chengfeng Cai.
Z.X. and D.X.Y. are supported by NKRDPC Grants No. 2017YFA0206203, No. 2018YFA0306001, NSFC-11574404, NSFG-2015A030313176, National Supercomputer Center in Guangzhou, and Leading Talent Program of Guangdong Special Projects. Z.Y. would
like to acknowledge the support
by Startup Grant (No. 74130-18841219) and NSFC-11904417.

\subsection{Appendix A: Derivation of
 the circuit complexity under the inner-product metric}

As mentioned in the main text, $|\psi_{k}\rangle$ is a superposition of
$|0\rangle$ and $c_{k}^{\dag}c_{-k}^{\dag}|0\rangle$. Therefore, under
the basis $(c_{k}^{\dag}c_{-k}^{\dag}|0\rangle,|0\rangle)$, $|\psi_{k}\rangle$
can be written as a two-component spinor,
\begin{eqnarray}
|\psi_{k}\rangle=\left(
                   \begin{array}{c}
                     i\sin(\theta_{k}/2) \\
                     \cos(\theta_{k}/2) \\
                   \end{array}
                 \right).
\end{eqnarray}
A two-component spinor can be mapped to a three-component vector in terms
of the Pauli matrices,
\begin{eqnarray}
\boldsymbol{\psi}_{k}=\langle \psi_{k}|\boldsymbol{\sigma}|\psi_{k}\rangle=-(0,\sin\theta_{k},\cos\theta_{k})^{T}.
\end{eqnarray}
Let us consider that the target state is infinitely close to the reference state, i.e.,
$|\psi^{R}\rangle=|\psi_{k}\rangle$, and $|\psi^{T}\rangle=|\psi_{k}+d\psi_{k}\rangle$,
or equivalently, $\boldsymbol{\psi}^{R}=\boldsymbol{\psi}_{k}$, and $\boldsymbol{\psi}^{T}=\boldsymbol{\psi}_{k}+d\boldsymbol{\psi}_{k}$.
The normalization requires $\boldsymbol{\psi}_{k}\cdot d\boldsymbol{\psi}_{k}=0$.
As the two states are infinitely close to each other, to leading order, the unitary operator which transforms the reference state to
the target state can be written as
\begin{eqnarray}
U=\mathbf{I}+i\mathbf{F}(\psi_{k})\cdot\boldsymbol{\sigma}+...
\end{eqnarray}
where $\mathbf{F}(\psi_{k})$ is an infinitely small parameter to be determined.
As $|\psi^{T}\rangle=U|\psi^{R}\rangle$ and $\boldsymbol{\psi}^{T}=\boldsymbol{\psi}_{k}+d\boldsymbol{\psi}_{k}=\langle \psi^{T}|\boldsymbol{\sigma}|\psi^{T}\rangle$, it is readily found that to leading order of $\mathbf{F}(\psi_{k})$.
\begin{eqnarray}
d\boldsymbol{\psi}_{k}&=&i\langle \psi_{k}|(\boldsymbol{\sigma}\mathbf{F}(\psi_{k})\cdot\boldsymbol{\sigma}
-\mathbf{F}(\psi_{k})\cdot\boldsymbol{\sigma}\boldsymbol{\sigma})|\psi_{k}\rangle\nonumber\\
&=&-2\langle \psi_{k}|\mathbf{F}(\psi_{k})\times \boldsymbol{\sigma}|\psi_{k}\rangle.
\end{eqnarray}
A solution of the above equation is
\begin{eqnarray}
\mathbf{F}(\psi_{k})=\frac{1}{2}(d\boldsymbol{\psi}_{k}\times\boldsymbol{\psi}_{k}+\alpha\boldsymbol{\psi}_{k}),
\end{eqnarray}
where $\alpha$ denotes  an infinitely small constant. Therefore,
to leading order, the unitary operator for an infinitely small change
of the state is given by
\begin{eqnarray}
U=\mathbf{I}+\frac{i}{2}(d\boldsymbol{\psi}_{k}\times\boldsymbol{\psi}_{k}+\alpha\boldsymbol{\psi}_{k})\cdot\boldsymbol{\sigma}.
\end{eqnarray}

Ignoring global phase, $U$ belongs to the SU(2) group. The standard inner-product metric on SU(2) is\cite{Brown2019com}
\begin{eqnarray}
ds^{2}&=&\frac{1}{2}\text{Tr}[dU^{\dag}dU]\nonumber\\
&=&\frac{(d\boldsymbol{\psi}_{k}\times\boldsymbol{\psi}_{k}+\alpha\boldsymbol{\psi}_{k})
\cdot(d\boldsymbol{\psi}_{k}\times\boldsymbol{\psi}_{k}+\alpha\boldsymbol{\psi}_{k})}{4}\nonumber\\
&=&\frac{d\boldsymbol{\psi}_{k}\cdot d\boldsymbol{\psi}_{k}+\alpha^{2}}{4},
\end{eqnarray}
which is minimized when $\alpha=0$. Accordingly, the inner-product metric is
$ds^{2}=d\boldsymbol{\psi}_{k}\cdot d\boldsymbol{\psi}_{k}/4=d\theta_{k}^{2}/4$.
Therefore, the circuit complexity, which corresponds to the geodesic length, is given by
\begin{eqnarray}
ds=|d\theta_{k}|/2=\arccos|\langle \psi_{k}+d \psi_{k}|\psi_{k}\rangle |.
\end{eqnarray}
Above we have used that
\begin{eqnarray}
|\psi_{k}+d\psi_{k}\rangle=\left(
                   \begin{array}{c}
                     i\sin((\theta_{k}+d\theta_{k})/2) \\
                     \cos((\theta_{k}+d\theta_{k})/2) \\
                   \end{array}
                 \right).
\end{eqnarray}
Performing a summation over $k$, we reach the expression of circuit complexity in Eq.(\ref{onecc}) of
the main text.

\subsection{Appendix B: The wave function of ground state}

We start with the Hamiltonian
$H=\frac{1}{2}\sum_{k}\psi_{k}^{\dag}H(k)\psi_{k}$ with
$\psi_{k}=(c_{k},c_{-k}^{\dag})^{T}$ and
\begin{eqnarray}
H(k)&=&(-t_{1}\cos k-t_{2}\cos 2k-\mu)\tau_{z}+(\Delta_{1}\sin k+\Delta_{2}\sin 2k)\tau_{y} \nonumber\\
&\equiv& d_{z}(k)\tau_{z}+d_{y}(k)\tau_{y}.\label{model}
\end{eqnarray}
By performing a standard Bogoliubov transformation, i.e.,
\begin{eqnarray}
c_{k}&=&\cos\frac{\theta_{k}}{2}\alpha_{k}+i\sin\frac{\theta_{k}}{2}\alpha_{-k}^{\dag},\nonumber\\
c_{-k}^{\dag}&=&i\sin\frac{\theta_{k}}{2}\alpha_{k}+\cos\frac{\theta_{k}}{2}\alpha_{-k}^{\dag},
\end{eqnarray}
where $\theta_{k}=\arctan (d_{y}(k)/d_{z}(k))$, the Hamiltonian
will be diagonalized as
\begin{eqnarray}
H&=&\frac{1}{2}\sum_{k}E_{k}(\alpha_{k}^{\dag}\alpha_{k}-\alpha_{-k}\alpha_{-k}^{\dag})\nonumber\\
&=&\sum_{k>0}E_{k}(\alpha_{k}^{\dag}\alpha_{k}+\alpha_{-k}^{\dag}\alpha_{-k})+...,
\end{eqnarray}
where $``...''$ stands for some unimportant constant, and $E_{k}=\sqrt{d_{y}^{2}(k)+d_{z}^{2}(k)}$.
The ground state satisfies $\alpha_{k}|\Omega\rangle=0$ for arbitrary $k$. One can find that
the solution is
\begin{eqnarray}
|\Omega\rangle=\prod_{k>0}|\psi_{k}\rangle=\prod_{k>0}
(\cos\frac{\theta_{k}}{2}+i\sin\frac{\theta_{k}}{2}c_{k}^{\dag}c_{-k}^{\dag})|0\rangle,\label{wave}
\end{eqnarray}
where $|0\rangle$ represents the vacuum, i.e., $c_{k}|0\rangle=0$. As $k$ is a good quantum number,
we can treat each $k$ independently. For each $k$, $|\psi_{k}\rangle$ is a superposition of
$|0\rangle$ and $c_{k}^{\dag}c_{-k}^{\dag}|0\rangle$. Let us focus on a specific $k$ and consider
that the reference state is given by $|\psi_{k}^{R}\rangle=(\cos\frac{\theta_{k}^{R}}{2}
+i\sin\frac{\theta_{k}^{R}}{2}c_{k}^{\dag}c_{-k}^{\dag})|0\rangle$, and the target state
is given by $|\psi_{k}^{T}\rangle=(\cos\frac{\theta_{k}^{T}}{2}
+i\sin\frac{\theta_{k}^{T}}{2}c_{k}^{\dag}c_{-k}^{\dag})|0\rangle$. The inner product of
the two states, which is also known as fidelity, is given by
\begin{eqnarray}
|\langle \psi_{k}^{T} | \psi_{k}^{R}\rangle|&=&|\cos(\theta_{k}^{T}/2)\cos(\theta_{k}^{R}/2)+\sin(\theta_{k}^{T}/2)\sin(\theta_{k}^{R}/2)|\nonumber\\
&=&|\cos((\theta_{k}^{T}-\theta_{k}^{R})/2)|\nonumber\\
&\equiv&|\cos(\Delta\theta_{k}/2)|.
\end{eqnarray}

\subsection{Appendix C: The growth rate of adiabatic circuit complexity}

Here we give a derivation of Eq.(\ref{agr}) in the main text. Consider a ground state
$|\Omega^{R}\rangle=\prod_{k>0}|\psi_{k}(\lambda)\rangle$ as the reference state,
where $\lambda$ refers to the parameter to vary, and consider its near neighbour
$|\Omega^{T}\rangle=\prod_{k>0}|\psi_{k}(\lambda+\delta\lambda)\rangle$ as the
target state, then according to Eq.(\ref{onecc}) in the main text, we have
\begin{eqnarray}
\delta C(\lambda)=\sum_{k>0}\arccos|\langle\psi_{k}(\lambda+\delta\lambda)|\psi_{k}(\lambda)\rangle |.
\end{eqnarray}
To characterize the growth rate of circuit complexity when  $\lambda$ is adiabatically varied away
from its reference state's value, a natural definition of the growth rate is
\begin{eqnarray}
\gamma(\lambda)\equiv \lim_{\delta\lambda\rightarrow 0^{+}}\frac{\delta C(\lambda)}{\delta\lambda}=\sum_{k>0}\lim_{\delta\lambda\rightarrow 0^{+}}\frac{\arccos|\langle\psi_{k}(\lambda+\delta\lambda)|\psi_{k}(\lambda)\rangle |}{\delta\lambda}.\quad
\end{eqnarray}
As $\delta\lambda$ is infinitely small, it's justified to make a Taylor expansion,
\begin{eqnarray}
\langle \psi_{k}(\lambda+\delta\lambda)|\psi_{k}(\lambda)\rangle
&=&\langle \psi_{k}(\lambda)|\psi_{k}(\lambda)\rangle+\langle \frac{d\psi_{k}(\lambda)}{d\lambda}|\psi_{k}(\lambda)\rangle\delta\lambda\nonumber\\
&&+\frac{1}{2}\langle \frac{d^{2}\psi_{k}(\lambda)}{d\lambda^{2}}|\psi_{k}(\lambda)\rangle(\delta\lambda)^{2}+....
\end{eqnarray}
According to the expression of $|\psi_{k}(\lambda)\rangle$, one can easily find that the linear-order term vanishes.
Therefore, up to second-order of $\delta\lambda$,
we have
\begin{eqnarray}
|\langle \psi_{k}(\lambda+\delta\lambda)|\psi_{k}(\lambda)\rangle|=1-\frac{1}{8}(\frac{d\theta_{k}}{d\lambda})^{2}(\delta\lambda)^{2}.
\end{eqnarray}
Accordingly, we have
\begin{eqnarray}
\delta C(\lambda)&=&\sum_{k>0}\arccos |\langle \psi_{k}(\lambda+\delta\lambda)|\psi_{k}(\lambda)\rangle|\nonumber\\
&=&\sum_{k>0}\arccos(1-\frac{1}{8}(\frac{d\theta_{k}(\lambda)}{d\lambda})^{2}(\delta\lambda)^{2})\nonumber\\
&=&\sum_{k>0}\sqrt{2}\sqrt{\frac{1}{8}(\frac{d\theta_{k}(\lambda)}{d\lambda})^{2}(\delta\lambda)^{2}}+...\nonumber\\
&=&\sum_{k>0}\frac{1}{2}|\frac{d\theta_{k}(\lambda)}{d\lambda}||\delta\lambda|,
\end{eqnarray}
and
\begin{eqnarray}
\gamma(\lambda)&=&\lim_{\delta\lambda\rightarrow 0^{+}}\frac{\delta C(\lambda)}{\delta\lambda}\nonumber\\
&=&\sum_{k>0}\frac{1}{2}|\frac{d\theta_{k}(\lambda)}{d\lambda}|.\label{growth}
\end{eqnarray}
To see that $\gamma(\lambda)$ is nonanalytic at critical points, we take a special case of the model in Eq.(\ref{model}) for illustration.
Concretely, we take $t_{2}=\Delta_{2}=0$ and $t_{1}=\Delta_{1}=1$. Accordingly, we have
\begin{eqnarray}
H(k)&=&(-\cos k-\mu)\tau_{z}+\sin k\tau_{y}.\label{kitaev}
\end{eqnarray}
For this Hamiltonian,  topological phase transitions  take place at
$\mu=\pm1$. As  only $\mu$ is a variable, we let $\lambda=\mu$, then we have
\begin{eqnarray}
\gamma(\mu)&=&\sum_{k>0}\frac{1}{2}|\frac{d\theta_{k}(\mu)}{d\mu}|\nonumber\\
&=&\frac{1}{2}\sum_{k>0}\frac{\sin k}{(\cos k+\mu)^{2}+\sin^{2}k}\nonumber\\
&=&\frac{L}{2}\int_{0}^{\pi}\frac{dk}{2\pi}\frac{\sin k}{(\cos k+\mu)^{2}+\sin^{2}k}\nonumber\\
&=&\frac{L}{4\pi}\int_{-1}^{1}\frac{dx}{1+\mu^{2}+2\mu x}\nonumber\\
&=&\frac{L}{4\pi\mu}\ln|\frac{1+\mu}{1-\mu}|,
\end{eqnarray}
where $L$ denotes the system size. One can see that $\gamma(\mu)$ exhibits
logarithmic divergence exactly at the two critical points $\mu=\pm1$.

\begin{figure}[t]
\centering
{\subfigure{\includegraphics[width=0.4\textwidth]{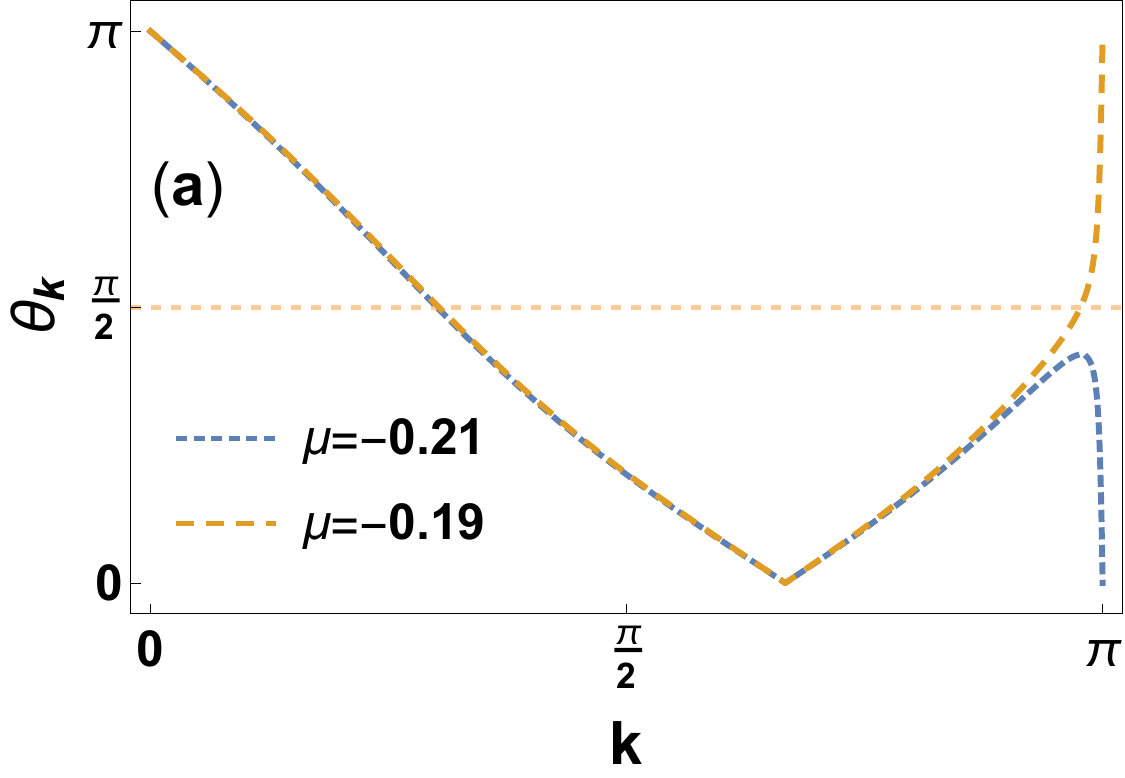}}}
{\subfigure{\includegraphics[width=0.4\textwidth]{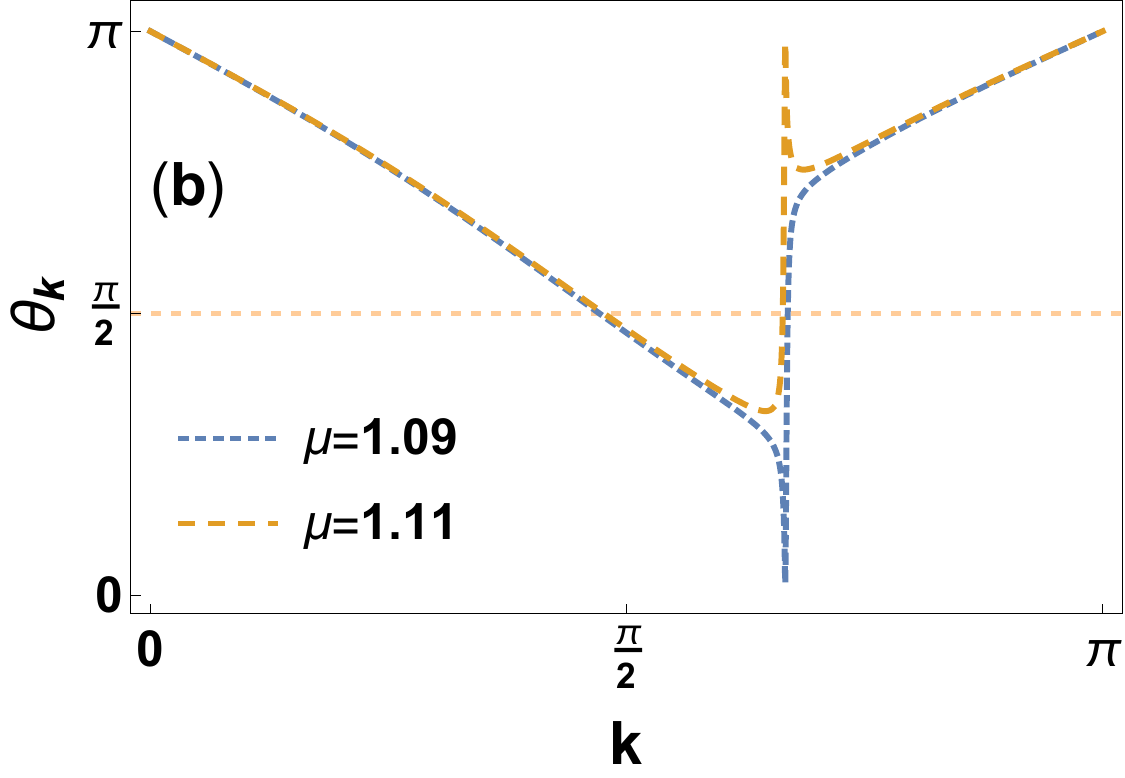}}}
\caption{ $\theta_{k}$-$k$ curves near critical points. Common parameters are $t_{1}=1$, $t_{2}=1.2$, and $\Delta_{1}=\Delta_{2}=1$. (a) Across the critical point $\mu_{c}=-0.2$, one can see that $\theta_{k}$ has a sudden jump at $k=\pi$.  (b) Across the critical point $\mu_{c}=1.1$,
$\theta_{k}$ has a sudden jump at $k=2\pi/3$. }\label{Sfig1}
\end{figure}

According to Eq.(\ref{growth}), one can understand that the presence of divergence in $\gamma(\mu)$ at a critical point
is because $\theta_{k}$ will undergo a dramatic change across a topological phase transition. To see this intuitively, we go back to
the more general model in Eq.(\ref{model})  and  plot $\theta_{k}$-$k$
curves near the critical points. As shown in Fig.\ref{Sfig1}, $\theta_{k}$ will undergo a sudden jump at some momenta
when the system goes across a topological phase transition. Such sudden jumps lead to the presence of divergence in $\gamma(\mu)$.

\subsection{Appendix D: Post-quench circuit complexity}

In the main text, we have considered a global quench of the Hamiltonian.
Concretely,  we consider that before quench, the system is described by $H_{i}$ and stays at its ground state $|\Omega^{i}\rangle$.
At $t=0$, the underlying Hamiltonian is suddenly quenched to $H_{f}$, and afterwards it keeps as $H_{f}$. For an isolated system,
the wave function follows a unitary evolution, i.e., $|\Omega(t)\rangle=e^{-iH_{f}t}|\Omega^{i}\rangle$ for $t>0$.
For each $k$, we have $|\psi_{k}(t)\rangle=e^{-iH_{f}(k)t}|\psi_{k}^{i}\rangle$. Thus, the post-quench circuit complexity
is given by $C(t)=\sum_{k>0}\arccos|\langle \psi_{k}(t)|\psi_{k}^{i}\rangle|$.

To obtain the concrete expression of $C(t)$, we need to know $|\psi_{k}(t)$. According
to the unitary evolution,
\begin{eqnarray}
|\psi_{k}(t)\rangle&=&e^{-iH_{f}(k)t}|\psi_{k}^{i}\rangle\nonumber\\
&=&e^{iE_{f}(k)t}\langle \psi_{k,-}^{f}|\psi_{k}^{i}\rangle|\psi_{k,-}^{f}\rangle
+e^{-iE_{f}(k)t}\langle \psi_{k,+}^{f}|\psi_{k}^{i}\rangle|\psi_{k,+}^{f}\rangle,\nonumber\\
\end{eqnarray}
where $|\psi_{k,\pm}^{f}\rangle$ are the eigenstates of $H_{f}(k)$,
with $|\psi_{k,+}^{f}\rangle$ and $|\psi_{k,-}^{f}\rangle$ corresponding to
eigenenergy $E_{f}(k)$ and $-E_{f}(k)$, respectively. Here
$E_{f}(k)=\sqrt{(d_{y}^{f}(k))^{2}+(d_{z}^{f}(k))^{2}}$.
According to Eq.(\ref{wave}), one has
\begin{eqnarray}
|\psi_{k,-}^{f}\rangle&=&(\cos\frac{\theta_{k}^{f}}{2}+i\sin\frac{\theta_{k}^{f}}{2}c_{k}^{\dag}c_{-k}^{\dag})|0\rangle,\nonumber\\
|\psi_{k,+}^{f}\rangle&=&(i\sin\frac{\theta_{k}^{f}}{2}+\cos\frac{\theta_{k}^{f}}{2}c_{k}^{\dag}c_{-k}^{\dag})|0\rangle,
\end{eqnarray}
then a short calculation reveals
\begin{eqnarray}
\langle \psi_{k}(t)|\psi_{k}^{i}\rangle
&=&e^{-iE_{f}(k)t}\langle\psi_{k,-}^{f}|\psi_{k}^{i}\rangle\langle\psi_{k}^{i}| \psi_{k,-}^{f}\rangle\nonumber\\
&&+e^{iE_{f}(k)t}\langle\psi_{k,+}^{f}|\psi_{k}^{i}\rangle\langle\psi_{k}^{i}| \psi_{k,+}^{f}\rangle\nonumber\\
&=&e^{-iE_{f}(k)t}\cos^{2}\frac{\Delta\theta_{k}}{2}+
e^{iE_{f}(k)t}\sin^{2}\frac{\Delta\theta_{k}}{2},
\end{eqnarray}
where $\Delta\theta_{k}=\theta_{k}^{f}-\theta_{k}^{i}$. Therefore,
\begin{eqnarray}
|\langle \psi_{k}(t)|\psi_{k}^{i}\rangle|&=&\sqrt{\cos^{2}(E_{f}(k)t)+\cos^{2}\Delta\theta_{k}\sin^{2}(E_{f}(k)t)}\nonumber\\
&=&\sqrt{1-\sin^{2}\Delta\theta_{k}\sin^{2}(E_{f}(k)t)}.
\end{eqnarray}
Accordingly, we obtain the expression for post-quench circuit complexity,
\begin{eqnarray}
C(t)=\sum_{k>0}\arccos\sqrt{1-\sin^{2}\Delta\theta_{k}\sin^{2} (E_{f}(k)t)}.\label{quenchcc}
\end{eqnarray}

\subsection{Appendix E: Growth rate of post-quench circuit complexity}

In the main text, we have shown that right after the quench, the post-quench circuit complexity grows
linearly with time. As the slope of a line can be determined by any two points on the line, thus the growth rate
of post-quench circuit complexity can be defined as
\begin{eqnarray}
\chi\equiv\lim_{\delta t\rightarrow 0^{+}}\frac{C(\delta t)-C(0)}{\delta t}=\lim_{\delta t\rightarrow 0^{+}}\frac{C(\delta t)}{\delta t}.
\end{eqnarray}
Above we have used  $C(0)=0$. According to Eq.(\ref{quenchcc}), we have
\begin{eqnarray}
\chi&=&\sum_{k>0} \lim_{\delta t\rightarrow 0^{+}}\frac{ \arccos\sqrt{1-\sin^{2}\Delta\theta_{k}\sin^{2} (E_{f}(k)\delta t)}}{\delta t}\nonumber\\
&=&\sum_{k>0}\lim_{\delta t\rightarrow 0^{+}}\frac{ \arccos(1-\frac{1}{2}\sin^{2}\Delta\theta_{k}(E_{f}(k)\delta t)^{2}
+\mathcal{O}((\delta t)^{3}))}{\delta t}\nonumber\\
&=&\sum_{k>0}\lim_{\delta t\rightarrow 0^{+}}\frac{\sqrt{2}\sqrt{\frac{1}{2}\sin^{2}\Delta\theta_{k}(E_{f}(k)\delta t)^{2}+\mathcal{O}((\delta t)^{3})}}{{\delta t}}\nonumber\\
&=&\sum_{k>0}|\sin\Delta\theta_{k}|E_{f}(k).
\end{eqnarray}
To see that $\chi$ is nonanalytic at the critical points, we again consider the special model in Eq.(\ref{kitaev}) for illustration.
Concretely, we consider
\begin{eqnarray}
H_{i}(k)&=&(-\cos k-\mu^{i})\tau_{z}+\sin k\tau_{y},\nonumber\\
H_{f}(k)&=&-\mu^{f}\tau_{z}.
\end{eqnarray}
Accordingly, we have
\begin{eqnarray}
\chi&=&\sum_{k>0}|\sin\Delta\theta_{k}|E_{f}(k)\nonumber\\
&=&\sum_{k>0}\frac{\sin k}{\sqrt{(\cos k+\mu^{i})^{2}+\sin^{2}k}}|\mu^{f}|\nonumber\\
&=&\frac{L|\mu^{f}|}{2\pi}\int^{\pi}_{0}dk\frac{\sin k }{\sqrt{(\cos k+\mu^{i})^{2}+\sin^{2}k}}\nonumber\\
&=&\frac{L|\mu^{f}|}{2\pi}\int_{-1}^{1}\frac{dx}{\sqrt{1+(\mu^{i}) ^{2}+2\mu^{i} x}}\nonumber\\
&=&\frac{L|\mu^{f}|}{2\pi}\frac{|1+\mu^{i}|-|1-\mu^{i}|}{2\mu^{i}},\nonumber\\
&=&\left\{
     \begin{array}{cc}
       \frac{L|\mu^{f}|}{2\pi}, & |\mu^{i}|\leq1, \\
       \frac{L|\mu^{f}|}{2\pi|\mu^{i}|}, & |\mu^{i}|>1. \\
     \end{array}
   \right.
\end{eqnarray}
The nonanalyticity of $\chi$ at the two critical points $\mu^{i}=\pm1$ is
obvious.

\bibliography{dirac}

\end{document}